\documentclass[amsmath,amssymb,11pt,superscriptaddress,reprint, preprintnumbers, notitlepage,aps,prl,twocolumn]{revtex4-1}
\pdfoutput=1 
\usepackage[utf8]{inputenc}
\usepackage[english]{babel}
\usepackage{amsmath}
\usepackage{graphicx}
\usepackage{dcolumn}
\usepackage{pbox}
\usepackage{amssymb}
\usepackage{epsfig}
\usepackage[dvipsnames]{xcolor}
\usepackage[normalem]{ulem}
\usepackage{slashed}
\usepackage{amssymb}
\usepackage{verbatim} 
\usepackage{ mathrsfs }
\usepackage{color}
\usepackage[font=small]{caption}
\usepackage[font=small]{subcaption}
\usepackage{url}
\definecolor{MyDarkBlue}{rgb}{0.1, 0.1, 0.8}
\definecolor{SBlue}{rgb}{0.2, 0.4, 0.7} 
\definecolor{MyLightBlue}{rgb}{0.22,0.51,0.9}
\definecolor{MyGreen}{rgb}{0.0, 0.5, 0.0}
\definecolor{BrickRed}{rgb}{0.8, 0.25, 0.33}
\RequirePackage{hyperref}
\hypersetup{colorlinks, citecolor=SBlue,linkcolor=MyDarkBlue, urlcolor=PineGreen}

\newcommand{\sigmav}{\ensuremath{{\langle \sigma v \rangle}}}
\newcommand{\meV}{\,\mathrm{meV}}
\newcommand{\eV}{\,\mathrm{eV}}

\newcommand{\MeV}{\,\mathrm{MeV}}
\newcommand{\GeV}{\,\mathrm{GeV}}

\makeatletter

\makeatletter
\renewcommand\@makecaption[2]{%
  \par
  \vskip\abovecaptionskip
  \begingroup
  
   \small\rmfamily
    \begingroup
     \samepage
     \flushing
     \let\footnote\@footnotemark@gobble
     \@make@capt@title{#1}{#2}\par
    \endgroup
  \endgroup
  \vskip\belowcaptionskip
}
\makeatother

\DeclareUnicodeCharacter{2212}{-}
\begin{document}
\title{\vspace{1cm}\Large 
Light  Neutrinophilic Dark Matter \\ from a  Scotogenic Model
}

\author{\bf Johannes Herms}
\email[E-mail:]{herms@mpi-hd.mpg.de}
\affiliation{Max-Planck-Institut f{\"u}r Kernphysik, Saupfercheckweg 1, 69117 Heidelberg, Germany}
\author{\bf Sudip Jana}
\email[E-mail:]{sudip.jana@mpi-hd.mpg.de}
\affiliation{Max-Planck-Institut f{\"u}r Kernphysik, Saupfercheckweg 1, 69117 Heidelberg, Germany}
\author{\bf Vishnu P.K.}
\email[E-mail:]{ vipadma@okstate.edu}
\affiliation{Department of Physics, Oklahoma State University, Stillwater, OK, 74078, USA}
\author{\bf Shaikh Saad}
\email[E-mail:]{ shaikh.saad@unibas.ch}
\affiliation{Department of Physics, University of Basel, Klingelbergstrasse 82, CH-4056 Basel, Switzerland}

\begin{abstract}
 We present a minimal sub-GeV thermal Dark Matter (DM) model where the DM primarily interacts with neutrinos and participates in neutrino mass generation through quantum loop corrections at one-loop level.
 We discuss the challenges in achieving this in the scotogenic framework and identify a viable variant.
 Due to minimality and the interplay between obtaining the correct DM relic abundance and neutrino oscillation data, the model predicts (i)  a massless lightest neutrino, (ii)  enhanced rate of $0\nu \beta \beta$  decay due to loop corrections involving light DM exchange, and (iii) testable lepton flavor-violating signal $\mu\to e\gamma$. Detecting monoenergetic neutrinos from DM annihilation in next-generation neutrino detectors offers a promising way to test this scenario. 
\end{abstract}

\maketitle

\textbf{\emph{Introduction}.--} 
The discovery of neutrino oscillations and the evidence for the existence of dark matter (DM) that constitutes about 26$\%$ of the Universe's energy density are two of the most compelling indications of particle physics beyond the Standard Model (BSM).
A captivating possibility lies in the potential explanation of both these phenomena through a unified framework of new physics.

Neutrino flavor oscillations imply the presence of non-zero neutrino masses and mixing, necessitating the construction of a model to describe neutrino masses.
Radiative neutrino mass models feature a lower scale of new physics associated with neutrino mass generation compared to tree-level scenarios, potentially leading to particles discoverable at current particle collider experiments.
The so-called ``scotogenic'' model~\cite{Tao:1996vb,Ma:2006km} is an elegant neutrino mass model along this line, where a symmetry that ensures loop-suppression of the neutrino mass is also responsible for the stability of dark matter.
To the best of our knowledge, the first proposal that linked neutrino mass and DM in this setup was made by Tao in 1996 \cite{Tao:1996vb}. However, it was not until 2006, when Ma independently introduced the radiative seesaw mechanism of neutrino mass and DM \cite{Ma:2006km} that it started getting attention in the community.
In the present work, we examine the possibility of sub-GeV thermal relic DM consistent with neutrino oscillation data in this framework.

The measured DM relic abundance can be related to particle physics properties through its production in the early Universe.
The most predictive dark matter production mechanism is the freeze-out of annihilation reactions.
It predicts small but significant couplings between the DM and the thermal bath particles and has motivated a large experimental program to look for such ``WIMP'' DM (see eg.~\cite{Arcadi:2017kky}). This has been so successful that many classes of DM models are now ruled out over large ranges of dark matter mass.
Dark matter coupled to neutrinos, however, has largely gone untested~\cite{Arguelles:2019ouk,Dutta:2022wdi}.
This is about to change, as next-generation large-volume neutrino detectors such as HyperKamiokande~\cite{Hyper-Kamiokande:2018ofw}, JUNO~\cite{JUNO:2015zny} and DUNE~\cite{DUNE:2015lol} go online.
They may be sensitive to the thermal relic value of the dark matter annihilation cross section in the sub-GeV WIMP regime, where the annihilation rate is largest.

This prospect motivates us to reconsider the simplest scotogenic model for neutrino masses and dark matter~\cite{Tao:1996vb, Ma:2006km}, adding one inert doublet and two right-handed neutrinos (RHNs) to the SM.
The possibility of a light scalar particle originating from the doublet~\cite{Herms:2022nhd} enables sub-GeV thermal relic DM in this scenario.
Following the analysis of~\cite{Boehm:2006mi}, we find that sub-GeV thermal freeze-out DM in the minimal scotogenic model is in tension with updated measurements of the active neutrino masses and cosmological bounds on the neutrino density in the Universe.

\begin{figure*}[thb]
\begin{center}
\includegraphics[width=0.75\textwidth]{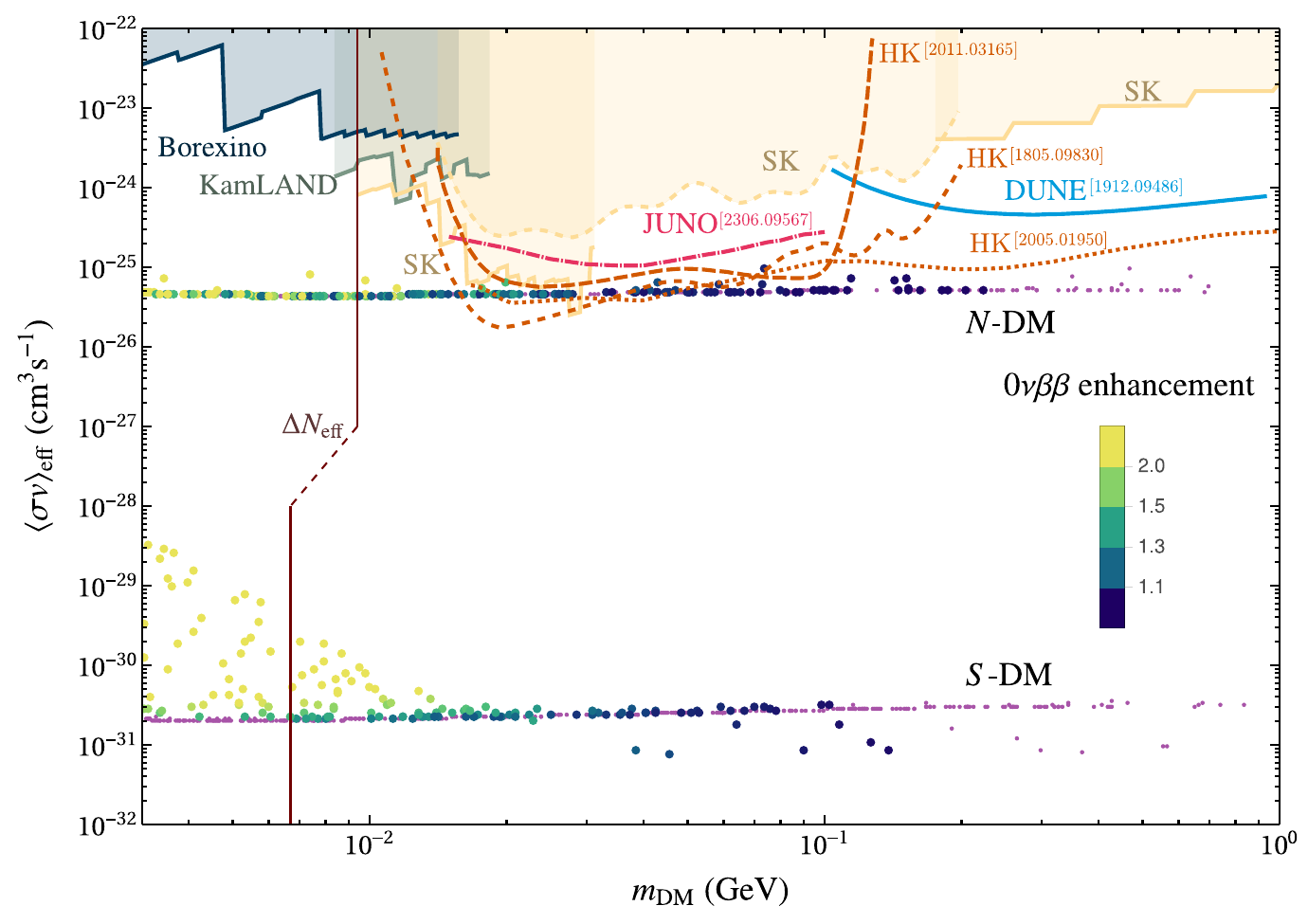}
\end{center}
\caption{
\label{fig:sigmav}
Predicted velocity-averaged annihilation cross section, taking $\langle v \rangle \sim 10^{-3}$ as typical for our Galaxy, with model parameters sampled from Fig.~\ref{fig:DMparameterSpace}.
Purple points result in excessive $\mu\to e\gamma$.
The color of the allowed points indicates the $0\nu\beta\beta$ enhancement (see text).
The plot also shows constraints from the Borexino~\cite{Borexino:2008gab}, KamLAND~\cite{KamLAND:2011bnd} and Super Kamiokande~\cite{Super-Kamiokande:2015qek} experiments and sensitivity forecasts for the planned Hyper-Kamiokande~\cite{Hyper-Kamiokande:2018ofw}, JUNO~\cite{JUNO:2023vyz} and DUNE~\cite{DUNE:2015lol} experiments, obtained by ~\cite{Arguelles:2019ouk} (line), ~\cite{Olivares-DelCampo:2018pdl} (dashed), ~\cite{Bell:2020rkw} (dotted), ~\cite{Asai:2020qlp} (long dashed), ~\cite{JUNO:2023vyz} (dot-dashed), assuming neutron tagging capability in Hyper-Kamiokande and differing runtimes of the experiments.
Scalar DM annihilation is $p$-wave suppressed. The red line indicates the present cosmological bound on the mass of neutrino-annihilating DM (as applicable to scalar and Dirac DM).
} 
\end{figure*}

In this work, we propose a viable and minimal version of the scotogenic model with sub-GeV DM that includes two inert scalar doublets and one Dirac singlet fermion.
After presenting the extended scalar sector, we obtain parameters that reproduce the observed neutrino masses and mixing while being in agreement with constraints on charged lepton flavor violation (cLFV).
We calculate the relic abundance and show that next-generation neutrino experiments will be sensitive to most of the (fermionic) DM parameter space.
These results are summarized in Fig.~\ref{fig:sigmav}.
To successfully replicate the DM relic density, this framework offers a light mediator, which along with the light DM, enhances the neutrinoless double beta decay ($0\nu\beta\beta$) rates due to loop contributions.

\medskip
\textbf{\emph{Challenges of light thermal DM in the minimal Scotogenic Model}--} 
Before presenting our proposed model, we will provide a concise overview of the minimal scotogenic model ~\cite{Tao:1996vb, Ma:2006km} and discuss the challenges involved in achieving light thermal dark matter within this framework.

The minimal scotogenic model ~\cite{Tao:1996vb,Ma:2006km} extends the particle content of the SM by one scalar doublet $\eta \sim (1,2,1/2)$ and two fermionic singlets $N_{1,2} \sim (1,1,0)$, all odd under a $\mathbb{Z}_2$ symmetry, whereas all SM fields are even.
The new Yukawa interaction
\begin{equation}
    -\mathcal{L} \supset Y_{i\alpha} \bar N_i \tilde \eta^\dagger l_\alpha  + \,\mathrm{h.c.},
\end{equation}
generates masses for the active neutrinos at the one-loop level~\cite{Ma:2006km,Merle:2015ica},
\begin{align}
    (\mathcal{M}_\nu)_{\alpha \beta} = \sum_i \frac{Y_{i \alpha} Y_{i \beta} M_i}{16 \pi^2} \bigg\{& \frac{m_R^2}{m_R^2 - M_i^2} \ln\frac{m_R^2}{M_i^2}
    \nonumber\\&
    -\frac{m_I^2}{m_I^2 - M_i^2} \ln\frac{m_I^2}{M_i^2}\bigg\}\,,
\end{align}
with $M_{1,2}$ the masses of the singlet fermions, and $m_{R,I}$ the masses of the real scalars $\sqrt{2} \Re\eta^0$ and $\sqrt{2} \Im\eta^0$.
The lightest of the $\mathbb{Z}_2$-odd particles is stable and will contribute to the DM in the Universe.
DM in the scotogenic model could live close to the electroweak scale, with its relic abundance determined by the Yukawa coupling, the gauge couplings of the scalars, or parameters of the scalar potential~\cite{Schmidt:2012yg,Abada:2018zra,LopezHonorez:2006gr}.

In this letter, we focus on the possibility that the lightest BSM particle (either an RHN or a neutral scalar) of the scotogenic model lives at the sub-GeV scale. To enable a sufficiently large annihilation cross section, whatever particle mediates the annihilation to neutrinos (a neutral scalar in case of fermionic DM or an RHN in case of scalar DM) is required to also be light~\cite{Okawa:2020jea}.
However, in order not to affect the known $Z$-decay width, at least one of the neutral scalars needs to be heavy, $m_R + m_I > m_Z$. As has been pointed out by~\cite{Boehm:2006mi}, DM is predicted to be at the MeV scale in this case by the following argument.
Taking for concreteness $m_{N_1},m_R \ll m_I$, the active neutrino mass induced by $N_1$ can be approximated by
\begin{equation}
    \label{eq:MnuN1}
    \mathcal{M}_\nu^{N_1} \simeq \frac{Y_1^2 m_{N_1}}{16 \pi^2} \log \left(\frac{m_I^2}{m_{N_1}^2}\right)\,,
\end{equation}
resulting in
\begin{equation}
    Y_1 \simeq 8.9\cdot 10^{-5} \times \sqrt{\frac{(\mathcal{M}_\nu^{N_1}/0.1\eV)}{(m_N/\GeV) \log(m_I/m_{N_1})}}\,.
\end{equation}
The relic abundance is determined by the thermally averaged annihilation cross section $\sigmav$~\cite{kolb_early_1990}
\begin{equation}
    \label{eq:relicAbundanceTextbook}
    \Omega h^2 \simeq \frac{x_f 1.07 \cdot 10^{9} \GeV^{-1}}{\sqrt{g_\mathrm{eff}(x_f)} m_\mathrm{Pl} \sigmav_s},
\end{equation}
with $x_f =m_\mathrm{DM}/T_{\mathrm{freeze-out}} \sim 18$ at thermal freeze-out.
The $s$-wave annihilation cross sections  read
\begin{align}
    (\sigma v)_{N_1 N_1\to \nu\nu}^{s-\mathrm{wave}} &= \frac{Y_1^4 m_{N_1}^2 (m_R^2 - m_I^2)^2}{128 \pi (m_{N_1}^2 + m_R^2)(m_{N_1}^2+m_I^2)},\\
    (\sigma v)_{RR\to \nu\nu}^{s-\mathrm{wave}} &= 2 \frac{Y_1^4 m_{N_1}^2}{8 \pi (m_{N_1}^2+m_R^2)^2} + \;\text{contrib.~from } N_2,
\end{align}
for $N_1$-DM and $\Re\eta$ DM, respectively. In the limit $m_R \ll m_I$, they differ only by a  numerical factor.
This leads to a prediction of the DM mass by demanding the observed relic abundance, $\Omega h^2 = 0.12$~\cite{Planck:2018vyg},
\begin{align}
    m_\mathrm{DM} \leq
    m_{N_1} =& \sqrt{\frac{\mathcal{M}_\nu^{N_1}}{0.1\eV} \frac{2}{1+\frac{m_R^2}{m_{N_1}^2}} \frac{\log\frac{110\GeV}{10\MeV}}{\log\frac{m_I}{m_{N_1}}}}
    \nonumber\\&
    \times
    \begin{cases}
        1.2 \MeV & R\mathrm{-DM} \\
        0.5 \MeV & N\mathrm{-DM}
    \end{cases}
    \label{eq:scotogenicDMbound}
\end{align}
for the  scalar and fermionic DM cases, respectively.
Larger values lead to DM overabundance.
On the other hand, $2\sigma$ bounds on excess radiation $\Delta N_\mathrm{eff}$ in the early Universe from BBN and CMB data restrict neutrino-annihilating DM to be heavier than $m_N \gtrsim 6 \MeV $ and $m_R \gtrsim 3 \MeV$~\cite{Sabti:2019mhn}.

The presence of the second RHN $N_2$ weakens this argument, as contributions to the active neutrio mass of form similar to Eq.~\ref{eq:MnuN1} can cancel the contribution from $N_1$ without significantly affecting the relic abundance.
In the absence of such tuned cancellation, sub-GeV DM in the minimal scotogenic model is in conflict with cosmological data.

In the next section, we propose a modified version of the scotogenic model that can accommodate light thermal dark matter.

\begin{figure}[thb!]
$$
\includegraphics[width=0.27\textwidth]{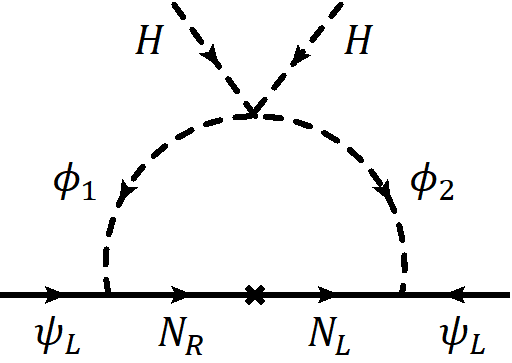}
$$
\caption{
\label{neutmass}
Representative Feynman diagram for neutrino mass. 
}
\end{figure}

\medskip
\textbf{\emph{Model}.--}
In the minimal scotogenic model, two RHNs and one inert scalar doublet are added to the SM to generate two nonzero masses for the active neutrinos.
Another minimal option is adding two inert scalars and one singlet fermion to the SM instead~\cite{Hehn:2012kz}.
We take this as a starting point to construct a model that breaks the direct link between $\mathcal{M}_\nu$ and $\sigmav$.
We add two scalar doublets $\phi_{1,2}$ and one singlet Dirac fermion $N_{L,R}$ to the particle content of the SM, all charged under a new $\mathcal{Z}_n$ symmetry.
The lightest of these dark scalars and the Dirac fermion may have masses below the GeV scale. They constitute a DM candidate and a light mediator to ensure sufficiently large annihilation rate~\cite{Okawa:2020jea}.
For $n>2$ \cite{Hagedorn:2018spx,Okada:2020oxh}, the neutrino mass diagram (Fig.~\ref{neutmass}) involves both of the scalar doublets, while DM annihilation may proceed through either. This breaks the direct link between DM annihilation and the scotogenic neutrino mass that was present in the minimal scotogenic model. 
Taking $n=3$ for concreteness, the new states have the following quantum numbers under the $SU(2)_L\times U(1)_Y\times \mathcal{Z}_3$ symmetry:
\begin{align}
&N_{R,L}\sim(1,0;\omega), \\
&\phi_1 = \begin{pmatrix} \phi_1^+  \\ \phi_1^0  \end{pmatrix}\sim (2,\dfrac{1}{2};\omega), \\ 
&\phi_2 = \begin{pmatrix} \phi_2^+  \\ \phi_2^0  \end{pmatrix}\sim (2,\dfrac{1}{2};\omega^2)\,,
\end{align}
with $\omega^3=1$.
The relevant Yukawa couplings of the model can be written as,
\begin{align}
-\mathcal{L}_{\rm Yuk} \supset &Y_1 \bar{\psi}_{L} \widetilde{\phi}_1 N_R + Y_2 \overline{\psi^c_L} \epsilon \phi_2 N_L 
\nonumber\\&+ M_N \bar{N}_{L}N_R + h.c. 
\label{Yuk1}
\end{align}
Here $\psi_L$ is the SM lepton doublet, $\epsilon$ is the 2-index Levi-Civita tensor, and $\widetilde{\phi}_i=\epsilon\phi^*_i$. 
The most general scalar potential consisting of the following terms:
\begin{align}
V(H,\phi_1,&\phi_2)=-\mu_{H}^2H^{\dagger}H + \mu_{\phi_1}^2\phi_1^{\dagger}\phi_1 + \mu_{\phi_2}^2\phi_2^{\dagger}\phi_2
\nonumber\\&
+\frac{\lambda_H}{2}(H^{\dagger}H )^2
+\frac{\lambda_{\phi_1}}{2}(\phi_1^{\dagger}\phi_1)^2
+\frac{\lambda_{\phi_2}}{2}(\phi_2^{\dagger}\phi_2)^2 
\nonumber\\&
+\lambda_{H\phi_1}(H^{\dagger}H)(\phi_1^{\dagger}\phi_1)
+\lambda_{H\phi_1}'(H^{\dagger}\phi_1)(\phi_1^{\dagger}H)
\nonumber\\&
+\lambda_{H\phi_2}(H^{\dagger}H)(\phi_2^{\dagger}\phi_2)
+\lambda_{H\phi_2}'(H^{\dagger}\phi_2)(\phi_2^{\dagger}H)
\nonumber\\&
+\left\{\lambda_{H\phi_1 \phi_2}(H^{\dagger}\phi_1)(H^{\dagger}\phi_2) + h.c.  \right\}
\nonumber\\&
+\left\{\lambda_{\phi_1 \phi_2 \phi_2}(\phi_1^{\dagger}\phi_2)(H^{\dagger}\phi_2)+  h.c.  \right\}
\nonumber\\&
+\left\{\lambda_{\phi_2 \phi_1 \phi_1}(\phi_2^{\dagger}\phi_1)(H^{\dagger}\phi_1) + h.c.  \right\}
\nonumber\\&
+\lambda_{\phi_1\phi_2}(\phi_1^{\dagger}\phi_1)(\phi_2^{\dagger}\phi_2)
+\lambda_{\phi_1\phi_2}'(\phi_1^{\dagger}\phi_2)(\phi_2^{\dagger}\phi_1). 
\end{align}
The Yukawa interactions given in Eq.~\eqref{Yuk1} and the $V\supset \lambda_{H\phi_1 \phi_2}(H^{\dagger}\phi_1)(H^{\dagger}\phi_2) + h.c.$ term in the scalar potential combinedly generate neutrino mass at the one-loop order, which is depicted  in Fig.~\ref{neutmass}.

The dark complex neutral scalars mix after electroweak symmetry breaking,
which is essential for generating non-zero neutrino masses.
The corresponding mass-squared matrix in the $\{\phi_1^0,\phi_2^{0*}\}$ basis is given by
\begin{align}
\mathcal{M}^2_{\phi^0}=
\begin{pmatrix}
    \mu^2_{\phi_1}+\frac{(\lambda_{H\phi_1}+\lambda'_{H\phi_1})}{2}v^2  &  \frac{\lambda_{H \phi_1 \phi_2} }{2}v^2\\
   \frac{\lambda_{H \phi_1 \phi_2} }{2}v^2 &  \mu^2_{\phi_2}+\frac{(\lambda_{H\phi_2}+\lambda'_{H\phi_2})}{2}v^2
\end{pmatrix}, \label{S0}
\end{align}
where $\langle H\rangle=v/\sqrt{2}$ with $v=246$ GeV. 
We denote the corresponding mass eigenstates  as $S_{1,2}$ (with masses $m_{S_{1,2}}$), which are related to $\{\phi_1^0,\phi_2^{0*}\}$   by
\begin{equation}\label{mix}
    \begin{pmatrix} S_1 \\ S_2^*\end{pmatrix}
    =\begin{pmatrix}
    \cos{\theta}  & -\sin{\theta} \\
	\sin{\theta} & \cos{\theta}
    \end{pmatrix}\begin{pmatrix} \phi_1^0\\ \phi_2^{0*} \end{pmatrix} \,.
\end{equation}
The mixing angle is given by
\begin{equation}
    \sin{2\theta}=\dfrac{\lambda_{H \phi_1 \phi_2} v^2 }{m^2_{S_2}-m^2_{S_1}}.
\end{equation}
On the contrary, the dark charged scalars $\phi_1^{\pm}$ and $\phi_2^{\pm}$ do not mix, and have the following masses:
\begin{equation}
    m^2_{\phi_1^{\pm}}=\mu^2_{\phi_1}+\frac{\lambda_{H\phi_1}}{2}v^2, \quad  m^2_{\phi_2^{\pm}}=\mu^2_{\phi_2}+\frac{\lambda_{H\phi_2}}{2}v^2.
\end{equation}
In the following,  before discussing its specific implications for DM phenomenology and neutrinoless double beta decay, we consider the requirements imposed on this model by scalar and lepton sector observables.

\medskip
\textbf{\emph{Scalar sector phenomenology}.--}
The requirement of a light mediator is the key to realizing sub-GeV DM in our scenario; this imposes stringent constraints on the scalar mass spectrum~\cite{Herms:2022nhd}, which we discuss in the following. 

The $Z$-boson can decay into a pair of light scalars $S_1$ at tree level, with amplitude proportional to $\cos{2\theta}$. Its measured decay width~\cite{Electroweak:2003ram,Tanabashi:2018oca} leaves little room for such additional invisible decay, fixing the neutral scalar mixing angle to $\theta\simeq\frac{\pi}{4}$.
To avoid decays $Z \to S_1 S_2$, we require $m_{S_2}\geq m_Z-m_{S_1}\simeq 90 \, \text{GeV}$.
The most stringent bound on the charged BSM scalars is imposed by the LEP experiment, $m_{\phi_{1,2}^{\pm}}\gtrsim 100 \, \text{GeV}$ \cite{Babu:2019mfe,Iguro:2022tmr}.
This leaves us with a mass hierarchy $m_{S_1}\ll m_{S_2},m_{\phi^{\pm}_{1,2}}\simeq \mathcal{O}(100)\,\text{GeV}$, with $\theta\simeq\frac{\pi}{4}$.

Mass splittings between members of the same electroweak multiplets affect electroweak precision observables, in particular the $T$-parameter~\cite{Peskin:1990zt,Peskin:1991sw}.
For computing this parameter, we follow Ref.~\cite{Grimus:2008nb}, and obtain 
 \begin{align} 
T  &=  \dfrac{1}{8\pi^2 \alpha_{\rm em}(M_Z) v^2 } \bigg\{ c_{\theta}^2  \mathcal{F}(m_{\phi_1^{+}}^2,m_{S_1}^2)
\nonumber\\&
+ s_{\theta}^2  \mathcal{F}(m_{\phi_2^{+}}^2,m_{S_1}^2) + s_{\theta}^2  \mathcal{F}(m_{\phi_1^{+}}^2,m_{S_2}^2)
\nonumber\\&
+c_{\theta}^2  \mathcal{F}(m_{\phi_2^{+}}^2,m_{S_2}^2)
    -s^2_{2\theta}\mathcal{F}(m_{S_1}^2,m_{S_2}^2) \bigg\} \,, \label{eq:T}
\end{align}
where  $\mathcal{F}$ is a symmetric function defined as
\begin{equation} \label{Fdef}
    \mathcal{F}(m_1^2,m_2^2) \  \equiv \  \frac{1}{2}(m_1^2+m_2^2) -\frac{m_1^2m_2^2}{m_1^2-m_2^2}\ln\left(\frac{m_1^2}{m_2^2}\right)\,.
\end{equation}
In the limit  $m_{S_1}\ll m_{S_2}= m_{\phi_1^{+}}= m_{\phi_2^{+}}$, the $T$-parameter takes a simple form,
\begin{align}
T=\frac{\cos^2{2\theta}\mathcal{F}(m_{S_1}^2,m_{S_2}^2)}{8\pi^2 \alpha_{\rm em}(M_Z) v^2} .    
\end{align}
This is automatically suppressed for the value $\theta\simeq\frac{\pi}{4}$ required to avoid invisible $Z$-decay.

Since in our scenario the neutral scalar $S_1$ is lighter than $m_h/2$, the SM Higgs $h$ can decay into $S_1^* S_1$  state via the following interaction term
\begin{equation}
    V\supset \frac{v}{2}(\lambda_{H \phi_1}+\lambda_{H \phi_2}+\lambda'_{H \phi_1}+\lambda'_{H \phi_2}-2\lambda_{H \phi_1\phi_2}) h (S_1^*S_1).
\end{equation}
The current limits on the branching ratio of the invisible decay of the SM Higgs ($\text{Br}_{\rm inv}$)  is $\text{Br}_{\rm inv}\leq 0.145$ at ATLAS \cite{ATLAS:2022yvh} and $\text{Br}_{\rm inv}\leq 0.18$ at CMS \cite{CMS:2022qva}. 
 To be consistent with these constraints,  we set the pertinent combination of quartic couplings to be small. This leads to the following relation
\begin{equation}
 \lambda_{H \phi_1}+\lambda_{H \phi_2} \simeq 2\lambda_{H \phi_1\phi_2}-(\lambda'_{H \phi_1}+\lambda'_{H \phi_2}). 
 \label{eq:inv}
\end{equation}
This has non-trivial implications for the radiative decay of the SM Higgs into two photons. The interaction between $h$ and the charged scalars $\{\phi_1^{\pm}  ,\phi_2^{\pm}\}$, 
\begin{equation}
    V\supset \lambda_{H \phi_1}v (h\phi_1^{+}\phi_1^{-})+\lambda_{H \phi_2}v (h\phi_2^{+}\phi_2^{-}).
\end{equation}
contributes to $h\to \gamma\gamma$ at one-loop level. Such contributions modify the Higgs signal strength into $\gamma \gamma$, 
\begin{equation}
    R_{\gamma\gamma}=\frac{\text{Br}(h\rightarrow \gamma\gamma)}{\text{Br}(h\rightarrow \gamma\gamma)_{\rm SM}}.
\end{equation}
For our scenario, the partial decay width of the $h\to \gamma\gamma$ process can be written as \cite{Okawa:2020jea, Babu:2018uik}  
\begin{align} 
&\Gamma(h\rightarrow \gamma\gamma)=\frac{\alpha^2 m_h^3}{256\pi^3 v^2}\left|\sum_f N_c^f Q_f^2 F_{1/2}(\tau_f) 
\nonumber\right.\\&\left. 
    + F_{1}(\tau_W) + \frac{(\lambda_{H\phi_1}+\lambda_{H\phi_2})v^2}{2 m_{S_2}^2} F_0(\tau_{S_2}) \right|^2.
    \label{eq:hgg}
\end{align}
Here $N_c^f$ and $Q_f$ stand for the color factor and the charge of the fermion $f$, respectively.   In Eq. \ref{eq:hgg}, the first two terms correspond to  the SM contributions, whereas the last term corresponds to the charged scalar contributions. The loop functions are defined as
\begin{align}
    F_{1/2} (\tau_i)&=-2\tau_i(1+(1-\tau_i)f(\tau_i)),\nonumber \\
    F_{1} (\tau_i)&=2+3\tau_i+3\tau_i(2-\tau_i)f(\tau_i), \nonumber \\
    F_{0} (\tau_i)&=\tau_i(1-\tau_if(\tau_i)),
\end{align}
with $\tau_i\equiv \dfrac{4m_{i}^2}{m_{h}^2}$ (where $m_i$ is the mass of the charged particle $i$ in the loop) and 
\begin{align}
    f(\tau_i)&= \begin{cases}
        \left[\sin^{-1}(\dfrac{1}{\sqrt{\tau}})\right]^2 & \tau \geq 1 \\
       -\frac{1}{4}\left[\ln\left(\dfrac{1+\sqrt{1-\tau}}{1-\sqrt{1+\tau}}\right)\right]^2 & \tau < 1 \,.
    \end{cases}
\end{align}
For the chosen mass spectrum $m_{S_1}\ll m_{S_2}\simeq m_{\phi_1^{+}}\simeq m_{\phi_2^{+}} $ with $\theta\simeq\frac{\pi}{4}$ and the scalar potential constraint Eq.~\ref{eq:inv}, we obtain
\begin{align}
    & \lambda_{H \phi_1}+\lambda_{H \phi_2} \simeq 2\lambda_{H \phi_1\phi_2}-(\lambda'_{H \phi_1}+\lambda'_{H \phi_2}) \simeq \frac{4m_{\phi^{\pm}}^2}{v^2},
\end{align}
where $m_{\phi^{\pm}}$ is the mass of the charged scalars $\phi_1^{\pm}$ and $\phi_2^{\pm}$. For $m_{\phi^{\pm}}\simeq \mathcal{O}(100)$ GeV, we find $R_{\gamma\gamma}\simeq 0.8$, which is consistent with the experimental results at $3\sigma$ level~\cite{ATLAS:2022tnm,CMS:2022wpo}.


\medskip
\textbf{\emph{Neutrino and cLFV Fit}.--} 
The one-loop neutrino mass diagram as shown in Fig.~\ref{neutmass} is calculable, and takes the following form~\cite{Ma:2006km,Merle:2015ica}: 
\begin{align}
\label{eq:mnu}
\mathcal{M}^\nu&=m_0\left(Y_1.Y_2^T+    Y_2.Y_1^T\right), 
\end{align}
with
\begin{align}
m_0&=\frac{m_{N}\sin 2\theta}{32\pi^2} 
\left\{\frac{m_{S_2}^2\ln\left( \frac{m_{S_2}^2}{m^2_N} \right) }{m_{S_2}^2-m_N^2}
-
\frac{m_{S_1}^2\ln\left( \frac{m_{S_1}^2}{m^2_N} \right) }{m_{S_1}^2-m_N^2}
\right\}.
\end{align}
%

To accommodate the large leptonic mixing angles, non-trivial flavor structures within $Y_{1,2}$ are required. This leads to cLFV processes $\ell_\alpha\to \ell_\beta\gamma$ mediated by the dark charged scalars, which can be expressed as~\cite{Vicente:2014wga},
\begin{align}
\label{eq:lfvBRformula}
&\frac{\mathrm{BR}\left(\ell_\alpha\to \ell_\beta\gamma\right)}{\mathrm{BR}\left(\ell_\alpha\to \ell_\beta\nu_\alpha\overline\nu_\beta\right)}
= \frac{3\alpha_\mathrm{EM}}{64\pi G^2_F}
\times 
\nonumber\\ &~~~~~~~~~~~~~~
\bigg\{
\left|\frac{Y_{1,\beta}^*Y_{1,\alpha}}{m^2_{\phi^+_1}}f\left( \frac{m_{N}^2}{m^2_{\phi^+_1}}\right)\right|^2 +\left(1\to 2 \right)  
\bigg\},
\end{align}
where
\begin{align}
f(x)=\frac{1-6x+3x^2+2x^3-6x^2\ln x}{6(1-x)^4} \,.
\end{align}
The most stringent limits on the Yukawa parameters in the model are set by $\mu\to e\gamma$.
The current upper bound is $Br(\mu\to e\gamma) < 4.2 \times 10^{−13}$ by the MEG collaboration~\cite{MEG:2016leq} and the next-generation experiment MEG-II~\cite{Baldini:2013ke} is set to improve sensitivity down to $Br(\mu\to e\gamma) < 6.4 \times 10^{−14}$.
Processes of the type $\ell_\alpha\to 3\ell_\beta$ are suppressed by an additional factor of 
$\alpha_\mathrm{EM}$.

We perform a fit to the neutrino masses and mixings, which is simultaneously consistent with cLFV and collider bounds mentioned above and results in MeV scale DM with the correct relic abundance.
For the simplicity of the analysis, we take all parameters to be real.  First, we fix the neutral scalar mixing angle to be $\theta=\pi/4$, and the mass spectrum is fixed to the following values: 
\begin{align}
&\left(M_N,m_{S_1}\right) = \left(20,100\right)\times 10^{-3}\; GeV,\\
&\left(m_{S_2},m_{\phi^+_1},m_{\phi^+_2}\right) = \left(110,110,110\right)\; GeV.
\end{align}
With these inputs, we obtain an excellent fit to the neutrino observables~\cite{NUFIT,Esteban:2020cvm}.
For normal neutrino mass ordering, the best fit parameters are
\begin{align}
&Y_1=10^{-2}\left(
\begin{array}{c}
 -0.24402 \\-1.76064 \\3 \\
\end{array}
\right),
\;
Y_2=10^{-7}\left(
\begin{array}{c}
-3.31953\\ 8.39950\\ -3.22219 \\
\end{array}
\right),
\end{align}
with resulting neutrino masses and mixing angles
\begin{align}
&\left(m_1,m_2,m_3\right)=\left(0, 0.86, 5.01\right)10^{-2}\; \eV,    
\\
&\left(\sin^2\theta_{12},\sin^2\theta_{13},\sin^2\theta_{23}\right)=\left(0.302, 0.571, 0.0223\right)\,,
\end{align}
and cLFV branching fractions
\begin{equation}
 BR(\mu\to \{e,\mu,\tau\} \gamma) = \{3.00,1.55,78.7\} \times 10^{-13}\,.
\end{equation}

For inverted neutrino mass ordering, smaller overall coupling strength $|Y_1|$ is required to avoid excessive $\mu\to e\gamma$. An example benchmark reads
\begin{align}
&\left(M_N,m_{S_1}\right) = \left(10,12.7\right)\times 10^{-3}\; GeV,
\end{align}
with Yukawa couplings
\begin{align}
&Y_1=10^{-3}\left(
\begin{array}{c}
 -1.35175\\-4.28086\\5.32767 \\
\end{array}
\right),
\;
Y_2=10^{-7}\left(
\begin{array}{c}
125.483\\-31.5768\\7.74377 \\
\end{array}
\right),
\end{align}
\begin{align}
&\left(m_1,m_2,m_3\right)=\left(49.23, 49.98, 0\right)\; \meV,    
\\
&\left(\sin^2\theta_{12},\sin^2\theta_{13},\sin^2\theta_{23}\right)=\left(0.304, 0.578, 0.0223\right). 
\end{align}
and corresponding cLFV branching fractions
\begin{equation}
 BR(\mu\to \{e,\mu,\tau\} \gamma) = \{5.44,1.50,14.6\} \times 10^{-15}\,.
\end{equation}
These benchmark points already include the relic density requirement, 
which is discussed in the next section.


\medskip
\textbf{\emph{DM phenomenology}.--} 
The DM phenomenology is determined by the two lightest dark particles $N, S_1$, the lighter of which is the DM, while the heavier one mediates the annihilation reactions that determine the relic abundance via the coupling
\begin{equation}
    -\mathcal{L} \supset
    \frac{1}{\sqrt{2}} \bar \nu (Y_1 P_R - Y_2 P_L) N S_1^* + \mathrm{h.c.}\,.
\end{equation}
The relevant cross sections read (cf.~\cite{Berlin:2014tja}) 
\begin{align}
    \sigma v_{s-\text{wave}}^{\bar N N \to \bar\nu\nu} &= \frac{m_N^2 (|Y_1|^2+|Y_2|^2)^2}{128 \pi (m_N^2 + m_{S_1}^2)^2}\\
    \sigma v_{s-\text{wave}}^{S_1^* S_1 \to \bar\nu\nu} &= \frac{m_N^2 |Y_1|^2 |Y_2|^2}{16 \pi (m_N^2 + m_{S_1}^2)^2}\\
    \sigma v_{p-\text{wave}}^{S_1^* S_1 \to \bar\nu\nu} &=
    \frac{m_{S_1}^2}{192 \pi \left(m_N^2+m_{S_1}^2\right)^4} \times \\
    \Big(\big(|Y_1|^4&+|Y_2|^4\big)\left(m_N^2+m_{S_1}^2\right)^2 -12 |Y_1|^2 |Y_2|^2 m_N^4\Big)
\end{align}
These only depend on the absolute value of the Yukawa vectors $Y_{1,2}$ and not on their flavor structure.
The neutrino mass (Eq.~\ref{eq:mnu})
predicts the combination
of both Yukawa couplings.
However, there are annihilation processes that depend on $Y_1$ and $Y_2$ separately. This allows for large annihilation rates in spite of small neutrino masses for $|Y_1| \gg |Y_2|$ (taking $|Y_1| > |Y_2|$ without loss of generality), accommodating the observed relic abundance in this extended scotogenic model.

Using Eq.~\ref{eq:relicAbundanceTextbook} with $x_f\sim 18$, $h_\mathrm{eff}(x_f)\sim 10$ (noting that $\sigmav_\mathrm{eff} = \frac{1}{2} \sigmav_{\overline{\mathrm{DM}}\,\mathrm{DM}\to\mathrm{SM}\,\mathrm{SM}}$ in the case of non-self-conjugate particles), we can determine the values of the couplings that lead to the observed relic abundance. 
In the fermionic DM case, $m_N < m_{S_1}$, we find
\begin{equation}
    |Y_1| \simeq 0.04 \sqrt{\frac{m_N^2+m_{S_1}^2}{m_N \GeV}} \,, 
\end{equation}
while in the scalar DM case, $m_{S_1} < m_N$, the hierarchical Yukawa couplings lead to $p$-wave dominance and require
\begin{equation}
    |Y_1| \simeq 0.07 \sqrt{\frac{m_N^2+m_{S_1}^2}{m_{S_1} \GeV}} \,. 
\end{equation}

\begin{figure}[thb]
$$
\includegraphics[width=0.4\textwidth]{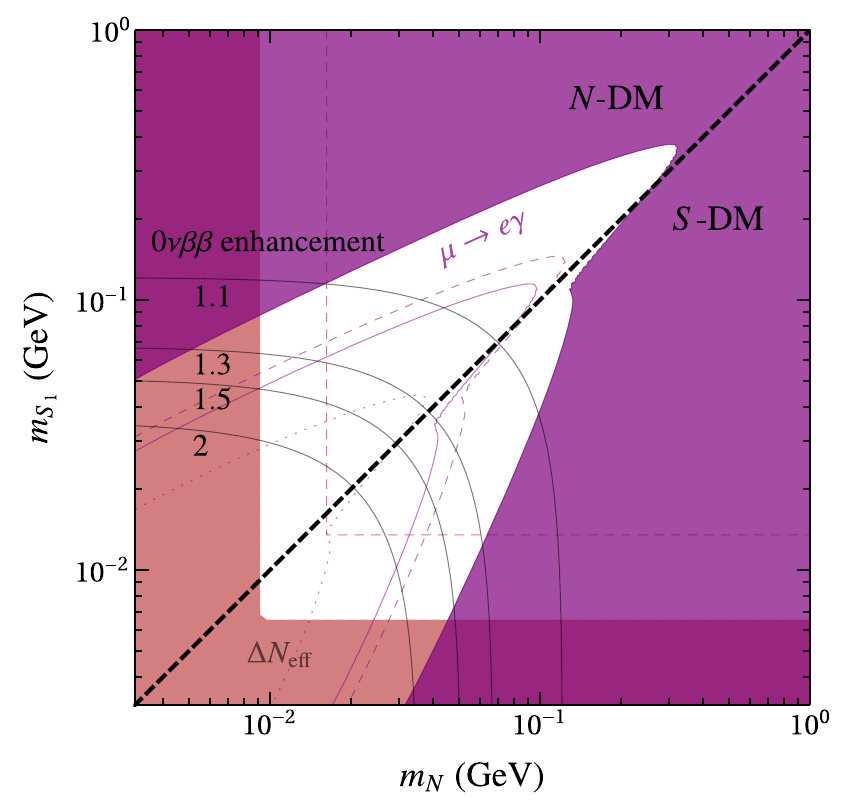}
$$
\caption{
Model parameter space. The Yukawa coupling strengths are chosen to reproduce neutrino masses and DM relic abundance. The white region of the plot is compatible with all constraints.
The red region is incompatible with bounds on the DM mass from CMB and BBN, the dashed line indicates future sensitivity~\cite{Sabti:2019mhn}.
The purple region can lead to excessive LFV for normal ordering of neutrino masses, assuming $m_{\phi^\pm}=110\GeV$. The constraint is stronger when demanding inverted ordering (line)  
and is weaker in case of larger charged scalar mass; future sensitivity~\cite{Baldini:2013ke} is shown as dashed (NO) and dotted (IO) lines.
The black contours indicate the $0\nu\beta\beta$ enhancement factor (\ref{eq:mbetabetalambda}), setting $p = 200 \MeV$ for concreteness.
}\label{fig:DMparameterSpace}
\end{figure}

In Fig.~\ref{fig:DMparameterSpace}, we show the parameter space of the model,
determining $|Y_1||Y_2|$ from the measured neutrino masses~\footnote{
The active neutrino masses can be written as
$m_{\{3,2\}}=2 m_0 |Y_1|\cdot|Y_2| \times \{\cos^2(\alpha/2),\sin^2(\alpha/2)\}$
with $Y_1 \cdot Y_2 = \cos \alpha \, e^{i \varphi} |Y_1|\cdot|Y_2|$. 
Reproducing the observed neutrino masses fixes the strength of the two Yukawa couplings
$|Y_1|\cdot|Y_2| = (m_3+m_2)/(2 m_0)$
}.
and $|Y_1|$ from the observed relic abundance (taking into account coannihilation processes~\cite{Griest:1990kh}), for different values of the DM and mediator masses, keeping all other BSM scalar masses at $110\GeV$ for concreteness. Taking a higher value of this mass, however, relaxes the cLFV constrains. 
The parameter space is constrained towards small masses by determinations of the expansion rate of the early Universe ($\Delta N_\mathrm{eff}$)~\cite{Sabti:2019mhn}.
Towards large masses, larger Yukawa couplings are required to maintain sufficient annihilation cross sections. We show conservative bounds on the strength of the Yukawa coupling $|Y_1|$ (Eq.~\ref{eq:lfvBRformula}) from $\mu\to e \gamma$ constraints in purple (see previous section).
Dashed lines indicate sensitivity forecasts for CMB-S4~\cite{Sabti:2019mhn} and MEG-II~\cite{Baldini:2013ke}.

\begin{figure*}[th!]
$$
\includegraphics[width=0.26\textwidth]{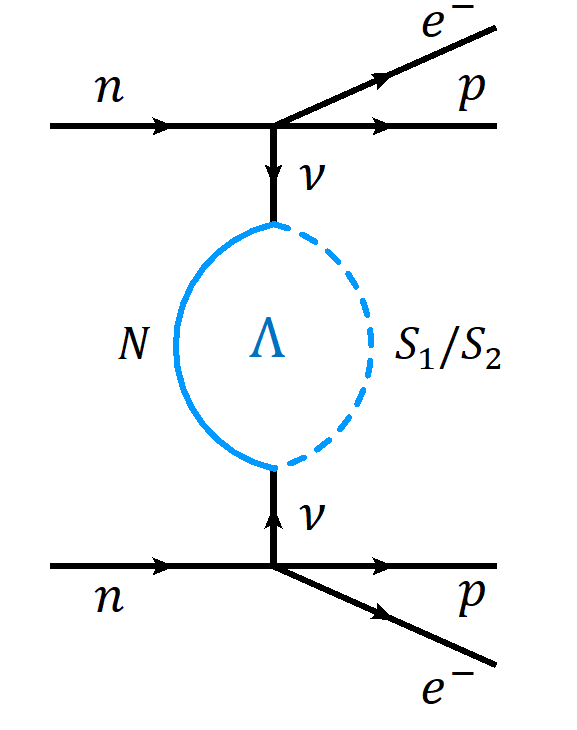}\hspace{0.72in}
\includegraphics[width=0.4\textwidth]{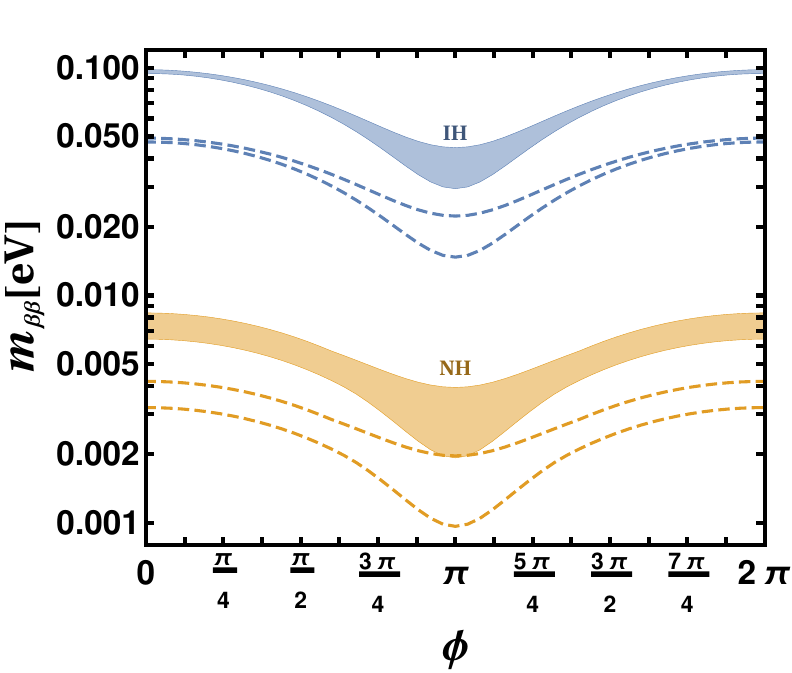}
$$
\caption{Left: Representative Feynman diagram for neutrino self-energy modified  $0\nu\beta\beta$ decay process; Right: Effective Majorana mass vs.  Majorana phase. The colored regions indicate the  effective Majorana mass for the scenario $\Lambda^2 \simeq p^2$. The dashed lines represent the  effective Majorana mass without the loop effect. 
} \label{fig:effectivemass}
\end{figure*}
\medskip
\textbf{\emph{Loop enhanced neutrinoless double beta decay}.--}
The neutrino self-energy can modify the rate of neutrinoless double decay in radiative neutrino mass models \cite{Rodejohann:2019quz} (see Fig.~\ref{fig:effectivemass}). This loop-induced contribution can be calculated by expanding neutrino self-energy $\Sigma(p)$ in terms of the external neutrino momentum, as follows
\begin{equation}
    \Sigma (p)=m_{\nu}\left(1+ \dfrac{p^2}{\Lambda^2}+\mathcal{O}(p^4)\right). 
\end{equation}
Here $m_{\nu}$ refers to the neutrino mass, while $\Lambda$ stands for the neutrino self-energy scale.
When $\Lambda^2 \simeq p^2$, this loop contribution becomes significant.
At the leading order,   the decay amplitude is modified as \cite{Rodejohann:2019quz}
\begin{equation}
    \label{eq:mbetabetalambda}
    \langle m_{\beta\beta} \rangle \longrightarrow \langle m_{\beta\beta} \rangle \left(1+\frac{p^2}{\Lambda^2}\right),
\end{equation}
where $\langle m_{\beta\beta} \rangle$ represents the $ee$-element of the neutrino mass matrix (effective neutrino mass). For computing $\Lambda$ for our scenario, we follow Ref. \cite{Rodejohann:2019quz}, and obtain
\begin{align}
&\Lambda^2=2 \frac{G[m^2_{S_2}]-G[m^2_{S_1}]}{H[m^2_{S_2}]-H[m^2_{S_1}]},
\end{align}
\begin{align}
&G[m^2_{S_i}]=\frac{m^2_{S_i}\ln{m^2_{S_i}}-m^2_{N}\ln{m^2_{N}}}{m^2_{S_i}-m^2_{N}},
\\
&H[m^2_{S_i}]=\frac{m^4_{N}-m^4_{S_i}+2m^2_{S_i}m^2_{N}\ln{\dfrac{m^2_{S_i}}{m^2_{N}}}}{(m^2_{S_i}-m^2_{N})^3}.
\end{align}
Notice that from the above equations, $\Lambda$ depends only on the masses of particles that participate in the neutrino mass generation.

The virtual neutrinos that mediate the decay process carry a momentum transfer $p\simeq \mathcal{O}(100)$ MeV \cite{Simkovic:2018hiq,Shimizu:2017qcy}.
On the other hand, in our scenario, the neutrino mass is generated at one-loop level with particles of a mass of similar energy scale. As a consequence of this, the $0\nu\beta\beta$ decay rate is significantly modified in our framework.  
To illustrate this loop-induced effect, in  Fig.~\ref{fig:effectivemass}, we plot the effective neutrino mass as a function of the Majorana phase $\phi$ when
$\Lambda^2 \simeq p^2$.  As noted earlier, the lightest neutrino is exactly massless in our scenario. This implies that for the normal (inverted) mass hierarchy, the effective neutrino mass depends on the relative  Majorana phase (one Majorana phase) only. The corresponding expressions for $\langle m_{\beta\beta} \rangle$  are given below:
\begin{align}
   \langle m_{\beta\beta}^{NH} \rangle &=\left|\sqrt{\Delta m^2_{21}}s_{12}^2 c_{13}^2 + \sqrt{\Delta m^2_{31}}s_{13}^2 e^{i\phi}  \right| \nonumber \\
   \langle m_{\beta\beta}^{IH} \rangle &=\left|\sqrt{|\Delta m^2_{32}| -\Delta m^2_{21}}c_{12}^2 c_{13}^2 + \sqrt{|\Delta m^2_{32}|}s_{12}^2 c_{13}^2 e^{i\phi}  \right|.
\end{align}
Here $\Delta m^2_{21}$ ($\Delta m^2_{3i}$) refers to the solar mass-splitting (atmospheric mass-splitting), $\theta_{ij}$ stands for the neutrino mixing angles (with $s^2_{ij}\equiv \sin^2{\theta_{ij}}$, $c^2_{ij}\equiv \cos^2{\theta_{ij}}$) and $\phi$ is the Majorana phase. We vary the fit values from {\tt NuFit5.2} \cite{Esteban:2020cvm} within the $3\sigma$ range. The colored regions indicate the effective neutrino mass for the  $\Lambda^2 \simeq p^2$ scenario, whereas the dashed lines correspond to the standard $0\nu\beta\beta$ decay process (with the lightest neutrino taken to be massless).

\medskip
\textbf{\emph{Conclusions}.--}
Annihilation into neutrinos is a simple thermal relic dark matter depletion mechanism that has so far mostly been beyond the experimental reach. However, upcoming neutrino detectors including Hyper-Kamiokande, DUNE, and JUNO, are set to become sensitive to this possibility for DM masses in the MeV range.
Motivated by these prospects, we investigated the scotogenic neutrino mass model for the possibility of light neutrino-annihilating dark matter. We find that a two-doublet one-Dirac-fermion version of the model enables sufficiently strong Dark Matter annihilation without resulting in excessive active neutrino masses.

The resulting annihilation cross sections are shown in Fig.~\ref{fig:sigmav}, alongside experimental bounds and sensitivity forecasts.
Reproducing the observed neutrino oscillation parameters determines the flavor structure of the new couplings of the model. This leads to inevitable contributions to $\mu\to e\gamma$ and constrains the parameter space towards large DM masses, complementing the DM indirect detection reach.
Remarkably, the requirement of a sub-GeV mediator involved in the process of dark matter annihilation can result in an intriguing amplification of neutrinoless double beta decay, which together with a monoenergetic neutrino signal from DM annihilation are distinctive predictions that establish connections between neutrino and dark matter phenomenology.

\vspace{0.1in}
\begin{acknowledgments}
{\textbf {\textit {Acknowledgments.--}}} We thank Carlos A. Argüelles for discussions and useful comments. The work of VPK is in part supported by US Department of Energy Grant Number DE-SC 0016013. For facilitating portions of this research, SJ and VPK wish to acknowledge the Center for Theoretical Underground Physics and Related Areas (CETUP*), The Institute for Underground Science at Sanford Underground Research Facility (SURF), and the South Dakota Science and Technology Authority for hospitality and financial support, as well as for providing a stimulating environment.
\end{acknowledgments}

\bibliographystyle{utphys}
\bibliography{reference}
\end{document}